\pdfoutput=1 

\documentclass{article}
\usepackage{float} 
\usepackage{natbib} 
\usepackage{amsmath} 
\usepackage{amsthm}
\usepackage{amsfonts} 
\usepackage{graphicx} 
\usepackage{multirow}
\usepackage[FIGTOPCAP,tight]{subfigure}

\usepackage{rotating}
\usepackage[all]{xy}
\input xy
\xyoption{all}

\newcommand{\bmbeta}{\boldsymbol{\beta}}

\newcommand{\bmpi}{\boldsymbol{\pi}}

\begin{document}

\title{A coupled hidden Markov model for disease interactions}
\author{Chris Sherlock$^1$,  \and Tatiana Xifara$^1$\footnote{Correspondence author: t.xifara@lancaster.ac.uk, Department of Mathematics and Statistics, Lancaster University, UK},\and
Sandra Telfer$^2$,
\and Mike Begon$^3$ \\
\\
\it
$^1$ Department of Mathematics and Statistics, Lancaster University, UK \\
\it $^2$ Institute of Biological and Environmental Sciences, University of Aberdeen, UK  \\
\it $^3$ Institute of Integrative Biology, University of Liverpool, UK
}
\date{5 March 2012}
\maketitle

\begin{abstract}
To investigate interactions between parasite species in a host, a population of field voles was studied longitudinally,
 with presence or absence of six different parasites measured repeatedly.
 Although trapping sessions were regular, a different set of voles was caught at each session leading
 to incomplete profiles for all subjects.  
We use a discrete-time hidden Markov model for each disease with transition probabilities dependent on covariates via a set
 of logistic regressions. For each disease the hidden states for each of 
the other diseases at a given time point form part of the covariate set for the Markov transition  probabilities from that time 
point. This allows us to gauge the influence of each parasite species on the transition probabilities for each of
 the other parasite species. Inference is performed via a Gibbs sampler, which cycles through each of the diseases, 
first using an adaptive Metropolis-Hastings step to sample from the conditional posterior of the covariate parameters for that particular
 disease given the hidden states for all other diseases and then sampling from the hidden states for that disease given the parameters.
We find evidence for interactions between several pairs of parasites and of an acquired immune response for two of the parasites.
\end{abstract}

{\bf Keywords}: Adaptive MCMC, Forward-Backward algorithm, Gibbs sampler, HMM, zoonosis.

\section{Introduction}
\subsection{Motivating problem}
\label{sec.motiv}
In natural populations, animals are likely to be infected by a variety of pathogens, either simultaneously or successively. Interactions between 
these pathogens, which can be synergistic or antagonistic, can affect infection biology (e.g. the intensity of one or both infections), host susceptibility 
to infection, or may impact on the host's morbidity or/and mortality. 
However, the biological processes involved are often too complex to allow clear-cut predictions regarding the outcome of such interactions. 
In order to explore potential interactions,
 a longitudinal study was undertaken by recording the sequences of infection events for different parasites in four spatially distinct populations of
 field voles ({\it Microtus agrestis}). The data are records of six pathogens: three species of {\it Bartonella} bacteria ({\it B. taylorii}, 
{\it B. grahamii}, {\it B. doshiae}), cowpox virus, the bacterium {\it Anaplasma phagocytophilum} and the protozoan {\it Babesia microti}. 
Aside from their intrinsic interest as a community of pathogens, {\it Bartonella}, {\it Anaplasma}, {\it Babesia} and cowpox virus infections may 
also be zoonotic: capable of being transmitted from animals to humans and causing disease.
 
As in most capture-mark-recapture studies, a different set of voles was caught at each session leading
to incomplete profiles for all subjects. 
The dataset therefore contains many missing observations; for example a profile
 for a given vole and a given disease from the first to last observation times for that
 vole might be $NPxxPNxP$, where $x$, $N$
 and $P$
 respectively indicate a missing observation, a negative response and a
 positive response. Inference on incomplete data in longitudinal and capture-recapture studies is a major problem; for examples see 
\cite{Daniels} and \cite{Pradel2005}. Previous analyses of our and 
related datasets (see \cite{Telfer2010} and \cite{Begon2009}) have
examined all pairs of observations for a given vole that occurred
exactly one lunar month apart and for which the first of the two
observations was an $N$. The influence of each covariate on the
probability of contracting a disease is then ascertained through logistic regression.   
In this paper we offer a more realistic model and a more powerful
analysis methodology for investigating the effects of previous infections for each disease on the other diseases. We use 
a hidden Markov model 
for each disease (Section \ref{sec: model}) and perform inference via
a Gibbs sampler; this allows us to use all of the dataset and to infer covariate effects on a given disease, 
even when these covariates are the (potentially missing or hidden) states of the other five diseases.  

 \subsection{Data}
\label{sec: data}
We analyse data collected between March 2005 and March 2007 from field voles in Kielder Forest, a man-made forest on the English-Scottish border.
 The voles were trapped at four grassy clear-cut sites within the forest, with each site at least $3.5$km from the nearest neighbouring site. 
Individuals were trapped within a 0.3ha live-trapping grid comprising 100 traps set at 5m intervals, with trapping taking place every 28 days
 from March to November, and every 56 days from November to March. \cite{Begon2009} provides further details of the study area and the trapping design. 

\begin{table}[!h]
\caption{\label{tbl:variables}Description of variables in the data set and their possible outcomes} 
 \centering
\fbox{%
\begin{tabular}{l l} 
\em{Variable} & \em{Description } \\ \hline 
  Tag & Unique number that identifies each vole \\
  Site & Identifier for the capture site ($4$ level factor)\\
  Sex & Male/Female \\
   Lm & Capture time point in whole lunar months (1 - 27, integer)\\
  Weight & Weight in grams rounded to the nearest 0.5$g$\\
  Sin & $\sin(2\pi \mbox{Lm}/13)$ \\
  Cos & $\cos(2\pi \mbox{Lm}/13)$\\ 
  \hline
  Tay &  {\it B. taylorii}, N(negative) or P(positive)\\
  Grah & {\it B. grahamii}, N(negative) or P(positive)\\
   Dosh & {\it B. doshiae}, N(negative) or P(positive)\\
   Cow & Cowpox, N(negative) or P(positive)\\   
   Ana & {\it Anaplasma}, N(negative) or P(positive)\\
   Bab & {\it Babesia}, N(negative) or P(positive)\\
\end{tabular}
}
\end{table}

Captured voles were marked with a unique identifying passive
transponder tag in order to be recognised in later captures. At each
capture, a $20 - 30\mu l$ blood sample was taken for pathogen
diagnostic tests. PCR assays were used to directly test for evidence
of infection with {\it Anaplasma phagocytophilum, Babesia microti} and
the three {\it Bartonella} spp. (see \cite{Courtney2004},
\cite{Bown2008} and \cite{Telfer2008}). Antibodies to cowpox virus
were detected by immunofluorescence assay (see \cite{Chantrey1999}).  A brief description
of the observed and derived variables is given in Table \ref{tbl:variables}.

\begin{table}[!h]
\caption{\label{tbl:missing} Frequency of missing values per vole as a function
  of the number of lunar months from the first capture time to the last capture time.}
\centering
\fbox{%
\begin{tabular}{c c c c c c}
\em{Lunar months from } &  \multicolumn{5}{c}{\em{Values missing}} \\ 
\cline{2-6} 
\em{first to last capture}& 0 & 1 & 2 & 3 & $\geq4$ \\
\hline
0 & 832 & - & - & -& - \\
1 & 275 & - & - & - & -\\
2 & 132 & 74 & - & - & -\\
3 & 75 & 55 & 15 & - & - \\
4 & 30 & 49 & 33 & 9 & - \\
5 & 21 & 24 & 34 & 5 & 3 \\
6 & 7 & 7 & 33 & 15 & 2 \\
7 & 1 & 4 & 25 & 16 & 9\\
$>$7&0&2&27&12&15\\
\hline
\end{tabular}
}
\end{table}

After some processing (described in detail in \cite{Xifara2012}) our dataset contains 4344 captures of 1841 voles.
Only voles that have been caught at least twice are directly informative about transition probabilities (see Section
\ref{sec: model}), although voles that have been captured only once
still contribute to inference for the initial distribution of each
hidden Markov chain (see Section \ref{sec: initial}).

The dataset contains a substantial fraction of missing data: almost half of the
voles are not captured at every lunar month between the first and last times they were observed.
Thus, even for many of the voles that were observed at least twice, not all of the covariates are available, either because the 
vole was not caught in a given lunar month, or sometimes because the vole was
caught but a given variable was not ascertained. Table
\ref{tbl:missing} shows the frequency of missing values derived from
the first cause. The number of additional missing values, where
it was not possible to ascertain the status of a particular disease,
despite the vole being captured, is given in Table
\ref{tbl:additional}. This table also shows the frequency of positive ($P$)
and negative ($N$) records for each disease.

\begin{table}
\caption{\label{tbl:additional} Summary information for the six
  diseases; number of missing values despite the vole being captured and numbers of
negative (N) and positive (P) responses.}
\centering
\fbox{%
\begin{tabular}{l c c c}
\em{Disease} & \em{\# additional} & \em{\# N} & \em{\# P} \\ 
& \em{missing values}  & &\\
\cline{1-4}
{\it B. doshiae} & 46 & 3583 & 715 \\
{\it B. grahamii} & 44 & 3468 & 832\\
{\it B. taylorii} & 32 & 3139 & 1173 \\
{\it Babesia} & 0 & 2354 & 1990 \\
Cowpox & 85 & 1408 & 2851 \\ 
{\it Anaplasma} & 6 &4107 & 231 \\
\end{tabular}
}
\end{table}

\subsection{Statistical challenges}
\label{sec:challenges}
We aim to investigate potential interactions between the six pathogens of the study. In particular, for each disease, $d$, we wish to evaluate
 the way in which the presence or absence of each of the other diseases (and perhaps further information such as whether or not any infection 
is in its first month) affects the probability of a vole contracting $d$. Additionally where applicable we are interested in how other diseases
 affect the probability of recovery from $d$.

We could model each disease as a two-state discrete-time Markov chain, where State 1 corresponds to no disease and
State 2 to presence of disease; however, this two-state model imposes a very
specific structure. For example the length of any infection is
geometrically distributed; however it might be that the probability of
remaining infected when a disease is in its first month (an acute
phase) is different to that in subsequent months (chronic phase).
It has also been found (e.g. \cite{Telfer2010}) that acute and chronic
phases of a disease $d_1$ can have different effects on the
probability of a vole contracting disease $d_2$. A two-state semi-Markov model
(see, for example, \cite{Guedon2003})
could account for the first effect, at the expense of extra
complexity, but not the second.
To adequately represent both the dynamics and influence of each disease
with minimal extra complexity,
therefore, in this analysis the dynamics of all
 but one of the diseases is modelled as a Markov chain with more
 than two states. 
Section \ref{sec: model}  details the model for each disease.

Only knowledge of the presence or absence
of the disease is available to us. In general, this equates to knowledge of a
subset of the state-space in which the true state must lie, but not to the
exact state of the chain. For example, for all but one disease, States
$2$ and $3$ both correspond to presence of the disease. 
In disease modelling, Hidden Markov Models (HMMs) arise when the
Markov model for disease
progression has a number of stages, or states, but these are not
directly observed (e.g. \cite{Guihennec2000},
\cite{Chadeau2010}). Often the relationship between the state of the
Markov chain and the
observation is stochastic, although in our case there is no
stochasticity involved, but the state of the
Markov chain is nonetheless hidden. Furthermore, observations are
only available to us when the vole has been captured. 
The forward-backward (FB) algorithm (see Section \ref{sec: FW-BW}) can be applied to
any discrete-time HMM with a
finite state-space and addresses both of these issues.

We consider $D=6$ diseases, and hence six interacting (or coupled) HMMs. It is possible to consider the coupled Markov chains for each disease
together as a single Markov chain on an extended state-space. 
In this case the likelihood
function is straightforward to evaluate using the forward-backward
algorithm (see e.g. \cite{Zucchini2009}) and a Bayesian analysis can
then be performed using MCMC. In our particular scenario the state-spaces have size $4,4,4,3,3,2$, which would lead
to an extended state-space of size $4^3\times 3^2 \times 2=1152$. Since the forward-backward algorithm applied to an HMM
with $n$ states takes $O(n^2)$ operations, a naive implementation of the algorithm applied to the extended state-space would be $O(1152^2)/O(3\times4^2+2\times
3^2+2^2)\approx O(18959)$ times less efficient computationally;
equivalently, $100000$ iterations of an algorithm which deals with
each chain separately would take the same CPU time as $5$ or $6$
iterations of the single-chain algorithm. In our specific scenario, but certainly not in generality, some of the transition
probabilities in each individual chain are zero, and (in our scenario) only $32768$ elements
of the extended transition matrix would be non-zero. The use of sparse
matrix routines could therefore reduce the efficiency ratio to
approximately $468$. Such a reduction in computational efficiency
would only be justified if fraction of missing data were very close to
$1$ so that the mixing of our Gibbs sampler would be extremely slow.

\cite{Pradel2005} analyses capture-recapture data using an HMM, and
incorporation of covariate information within this framework via an
appropriate link function is straightforward (see \cite{Lachlish2011},
\cite{Zucchini2009} (Section 8.5.2)). However the methodology does not allow the use of multiple HMMs nor, therefore, can it use the state of each HMM as a covariate for the other HMMs. We require six HMMs (one for each disease) and we wish to use covariate information such as the time of year and weight of the vole.  Furthermore we wish the covariate set for each disease to include the states of the HMMs for the other diseases. For each disease, $d$,  we will represent the probability of each possible state change through a logistic regression. However some of the covariates, the states of the other HMMs, are unknown. Our solution is a Gibbs sampler which employs the forward-backward algorithm and adaptive random walk Metropolis steps to jointly sample from the true posterior distribution of all of the HMMs and the covariate parameters.

\subsection{Outline}
 The remainder of this paper is organised as follows. Section 2
 describes the model which was used for each disease, gives its
 likelihood function, and outlines the imputation of missing weight
 values and the other fixed covariate values. 
 The Markov chain Monte Carlo 
algorithm is described in Section 3 and we present our results, including the sensitivity study, in Section 4. The paper concludes with a discussion.

\section{Modelling the hidden and missing data}

\subsection{Hidden Markov models and notation}
\label{sec: model}

Hidden Markov models (HMM) are used when observations are influenced
 by a Markov process but the state of the Markov process itself cannot
 be determined exactly from the observations. Usually the relationship
 between the Markov process and the observation process 
is stochastic, but (as in our application) this need not
 be the case. For various examples and applications
 of HMMs see, for example, \cite{Zucchini2009}. Purely to simplify our
 subscript notation
 we consider each vole to have been first observed at a local (to the
 vole) time of $1$ and
 last observed at (local) time $T$. For disease $d$ ($d=1,\ldots,D$), let the state space for the Markov
 chain be $\mathbb{S}^{[d]}$ and the state space for the observation process be
$\mathbb{Y}:=\{N,P\}$. 
For a  given vole and for disease $d$, the unobserved Markov chain and
the observations are respectively
\[
{\bf X}^{[d]}:=
 \left(X_{1}^{[d]},X_{2}^{[d]},\ldots,X_{T}^{[d]}\right)\in
 \left(\mathbb{S}^{[d]}\right)^T
~\mbox{and}~
{\bf Y}^{[d]}:=\left(Y_{1}^{[d]},Y_{2}^{[d]},\ldots,Y_{T}^{[d]}\right)\in
\left(\mathbb{Y}\cup\{\mbox{missing}\}\right)^T.
\]
Note that $Y_{t}^{[d]}$ is conditionally independent of $X_{1}^{[d]},\ldots,X_{t-1}^{[d]},X_{t+1}^{[d]},\ldots,X_{T}^{[d]}$ given
$X_{t}^{[d]}$. 
The observed process $\{Y^{[d]}\} \in \mathbb{Y}$ is
related to the state of the hidden process $\{ X^{[d]}\}$ by a
likelihood vector, ${\bf l}_t^{[d]}$, which
has elements: 
\[
l_{t,i}^{[d]}=
\left\{
\begin{array}{ll}
\mathbb{P}\left(Y^{[d]}=y_t^{[d]}|x_t^{[d]}=i\right) &\mbox{if } y_t^{[d]}\in \{N,P\}\\
1&\mbox{if }y_t^{[d]}=\mbox{missing}.
\end{array}
\right.
\]
This vector is defined for each disease in 
Section \ref{sec: HMMdetail}.

We take the discrete time interval of each Markov chain to be one
lunar month. Since trapping sessions in winter took
place every two lunar months (see Section \ref{sec: data}) this inevitably leads to missing
observations for any vole caught several times over the winter, even
if it is caught at every trapping session.
For each unknown transition probability (see Section \ref{sec: HMMdetail}) we have a logistic regression model;
for example the probability, $p_{11,t}^{[d]}$, that a given vole will be
in state 1 (disease $d$ absent) at time
$t+1$ given that it is in state 1 at time $t$ is given by
$$logit(p_{11,t}^{[d]})=\left({\bf z}_{t}^{[d]}\right)^T\boldsymbol{\beta}_{11}^{[d]}.$$
Here ${\bf z}^{[d]}_{t}$ is the vector of covariates at time $t$, which
for all models includes the states of the other diseases at time $t$,
${\bf x}_t^{[-d]}$, as well as a deterministic covariate vector, ${\bf
  z}^{*[d]}_t$. This deterministic vector was chosen via forward fitting of logistic regression
models that were very similar to those of \cite{Telfer2010} and
\cite{Begon2009} (see Section
\ref{sec.motiv}). However, whereas \cite{Telfer2010} and
\cite{Begon2009} allow both the current covariates and
covariates one lunar month into the future to influence the response
one lunar month into the future, we only allow the current covariates to influence the
future response; further details are available in \cite{Xifara2012}. For all diseases the deterministic vector consists of 
 a time trend (lunar month, \texttt{Lm}, as a continuous covariate), a
seasonal cycle in the form of \texttt{sin} and \texttt{cos}, and
\texttt{sex}, \texttt{weight}, and \texttt{site}. The
covariate vector for cowpox
also includes a different trend with lunar month for each site, and
for all other diseases it allows for a different seasonal cycle for
each \texttt{sex} (see Table \ref{tbl:variables} for detailed covariate descriptions).

 \begin{figure}[!ht]
\begin{small}
  \xymatrix{  & & & *+<12pt>[o][F-]{\boldsymbol{\beta}^{[1]}} \ar@/_2pc/[dd] \ar@/^/[ddr] \ar@/^/[ddrr]  \ar@/^3pc/[ddrrr]& & & &  \\
  &   & *+<8pt>[F]{\txt{$y_{1}^{[1]}$}} &
    *+<8pt>[F]{\txt{$y_{2}^{[1]}$}} &
    & *+<8pt>[F]{\txt{$y_{4}^{[1]}$}}& *+<8pt>[F]{\txt{$y_{5}^{[1]}$}} & \\
    & *+<12pt>[o][F-]{\bmpi^{[1]}} \ar[r]&
    *+<10pt>[o][F-]{\txt{$X_{1}^{[1]}$}}\ar[r] \ar[u]_{{\bf l}^{[1]}_1} \ar[dr] |!{[d];[r]}\hole&
    *+<12pt>[o][F-]{\txt{$X_{2}^{[1]}$}}\ar[r] \ar[u]_{{\bf l}^{[1]}_2} \ar[dr] |!{[d];[r]}\hole&
    *+<10pt>[o][F-]{\txt{$X_{3}^{[1]}$}}\ar[r] \ar[dr] |!{[d];[r]}\hole& *+<10pt>[o][F-]{\txt{$X_{4}^{[1]}$}}\ar[u]_{{\bf l}^{[1]}_4} \ar[r] \ar[dr] |!{[d];[r]}\hole&*+<10pt>[o][F-]{\txt{$X_{5}^{[1]}$}}\ar[u]_{{\bf l}^{[1]}_5} &
     \\
    & *+<11pt>[o][F-]{\bmpi^{[2]}} \ar[r]&
    *+<10pt>[o][F-]{\txt{$X_{1}^{[2]}$}}\ar[r] \ar[ur] \ar[d]^{{\bf l}^{[2]}_1}&
    *+<12pt>[o][F-]{\txt{$X_{2}^{[2]}$}}\ar[r] \ar[ur] &
    *+<10pt>[o][F-]{\txt{$X_{3}^{[2]}$}}\ar[r] \ar[ur] & *+<10pt>[o][F-]{\txt{$X_{4}^{[2]}$}} \ar[r] \ar[d]^{{\bf l}^{[2]}_4}  \ar[ur]&  *+<10pt>[o][F-]{\txt{$X_{5}^{[2]}$}} \ar[d]^{{\bf l}^{[2]}_5}&
    \\
&    & *+<8pt>[F]{\txt{$y_{1}^{[2]}$}} & &  & *+<8pt>[F]{\txt{$y_{4}^{[2]}$}}& *+<8pt>[F]{\txt{$y_{5}^{[2]}$}}&   \\
     & & & *+<12pt>[o][F-]{\boldsymbol{\beta}^{[2]}} \ar@/^/[uu] \ar@/_/[uur] \ar@/_/[uurr] \ar@/_3pc/[uurrr]& & &  & }
\end{small} 
  \caption{Directed graph of a realisation of two parallel hidden
    Markov models; where $y_{t}^{[d]},$ $t=1,\ldots,5,$ where present, are the
    observed values of disease $d$ ($d=1,2$) and $X_{t}^{[d]},$
    $t=1,\ldots,5,$ are the states of the hidden Markov chain for
    $d$, arising from (unknown) initial distribution $\bmpi^{[d]}$. The nodes from $\boldsymbol{\beta}^{[d]}$ reflect dependence
    of the transition matrix on the (unknown) covariate parameters. To simplify 
    the presentation and focus on the coupled HMM we omit the
    deterministic covariate vectors ${\bf z}^{*[d]}_t$.}
   \label{fig: HMM}
\end{figure}
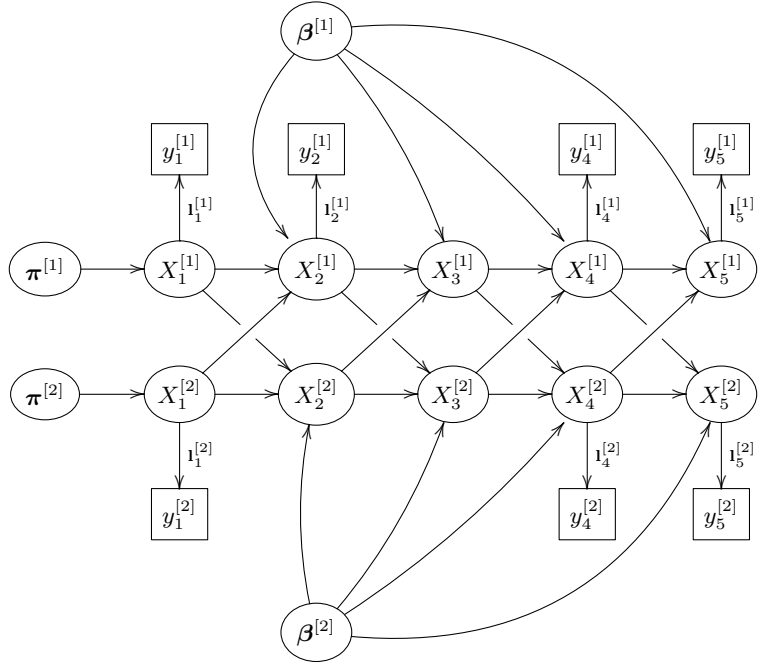

We denote the transition probability matrix from time $t$ to $t+1$ for
disease $d$ by $
{\it P}^{[d]}\left(\bmbeta^{[d]},{\bf x}_t^{[-d]},{\bf
    z}_t^{[d]}\right)$, i.e. ${\it P}^{[d]}_{i,j}\left(\bmbeta^{[d]},{\bf x}_t^{[-d]},{\bf z}_t^{[d]}\right)= Pr\left(X_{t+1}^{[d]}=j|X_{t}^{[d]}=i\right)$, and let the initial distribution for the hidden chain
be $\bmpi^{[d]}$. Figure \ref{fig: HMM} depicts a simplification of our scenario, where
there are just two diseases. Note that the states of all chains at
time $t+1$, $X_{t+1}^{[1]},\ldots,X_{t+1}^{[D]}$, are independent
conditional on the states of all chains at time $t$.

\subsection{Likelihood function}
\label{sec: likelihood} 

We now provide full detail of the likelihood for a given vole. The
likelihood for the data is simply the product of these likelihoods
over all 1841 voles. 
Let ${\boldsymbol \beta}^{[1:D]}:=({\boldsymbol
  \beta}^{[1]},{\boldsymbol \beta}^{[2]},\ldots,{\boldsymbol
  \beta}^{[D]})$, ${\bf y}^{[1:D]}=({\bf y}^{[1]},{\bf
  y}^{[2]},\ldots,{\bf y}^{[D]})$ and ${\bf x}^{[1:D]}=({\bf
  x}^{[1]},{\bf x}^{[2]},\ldots,{\bf x}^{[D]})$. 
The conditional independence structure leads to a complete data likelihood of
\[
L\left({\bf y}^{[1:D]},{\bf x}^{[1:D]};\bmbeta^{[1:D]},\bmpi^{[1:D]}\right)= 
 \prod_{d=1}^D \pi^{[d]}_{x_1^{[d]}}l_{1,x_1^{[d]}}^{[d]}
 \prod_{t=1}^{T-1}
 P^{[d]}_{x_t^{[d]},x_{t+1}^{[d]}}\left(\bmbeta^{[d]},{\bf x}_t^{[-d]},{\bf
     z}_{t}^{[d]}\right)l_{t+1,x_t^{[d]}}^{[d]}.
\]
The observed data likelihood for the vole is then
\begin{equation}
\label{eqn.likelihood}
L\left({\bf y}^{[1:D]};\bmbeta^{[1:D]},\bmpi^{[1:D]}\right)
=\sum_{{\bf x}^{[1:D]}\in \mathbb{S}_*^T}L\left({\bf y}^{[1:D]},{\bf x}^{[1:D]};\bmbeta^{[1:D]},\bmpi^{[1:D]}\right),
\end{equation}
where $\mathbb{S}_*:=\mathbb{S}^{[1]}\times \dots \times \mathbb{S}^{[D]}$.
For a single chain the summation over the hidden states
can be written as a matrix product; this simplification is not
possible for coupled chains as the transition matrix for each disease
depends on the state of each of the other diseases.  
Our Bayesian analysis requires prior distributions for
$\bmbeta^{[1:D]}$ and $\bmpi^{[1:D]}$, which are detailed in Section
\ref{sec: prior}. The product of the observed data likelihoods over
all voles multiplied by the prior distribution for $\bmbeta^{[1:D]}$
and $\bmpi^{[1:D]}$ gives, up to a constant of proportionality, the
joint posterior for $\bmbeta^{[1:D]}$, $\bmpi^{[1:D]}$ and ${\bf x}^{[1:D]}$.

\subsection{The forward-backward algorithm}
\label{sec: FW-BW}
The forward-backward algorithm developed by Baum {\it et al.} (1970)
(see also \cite{Zucchini2009}, \cite{Scott2002}, \cite{Rabiner1989}
and \cite{Chib1996}, for example) may be applied to any
discrete-time hidden Markov model with a finite state-space, and provides us with two useful tools. The {\it
  forward recursion} is a computationally efficient
algorithm for calculating the likelihood of the observed
data, while the {\it backward recursion}  provides us
with the distribution of each hidden state, $X_t$, given the state at
the next time point, $x_{t+1}$, and all of the observations. Both will form part
of the Gibbs sampling scheme that will be described in detail in
Section \ref{sec: gibbs}. 

\subsection{Other missing covariates}
\label{sec: fixed}
As mentioned in Section \ref{sec:challenges}, for many voles not all of the covariates are available. For a given vole, the covariates \texttt{sex}, \texttt{site},
and \texttt{Lm} clearly carry over to the missing records.
The unobserved disease states will be treated dynamically and will be sampled from the conditional distribution as part of the Gibbs sampling scheme 
(see Section \ref{sec: gibbs}).  Such sampling could
perhaps also be performed for \texttt{weight}. However here we adopt a simpler approach whereby
each missing weight value  
 is imputed once via linear interpolation between the two
nearest observed values for that vole. 
The robustness of inference to other sensible imputed weight
values obtained by using a growth model is investigated in Section \ref{sec: sensitivity}.

\subsection{Details of the Markov models for individual diseases}
\label{sec: HMMdetail}
The remainder of this section gives a brief description of each
disease in the study and describes the hidden Markov model that is used to
model it. All transition probabilities are time dependent since
some of the
covariates are time-dependent; however for ease of notation we drop any explicit reference to
this time dependence. A more detailed description of the host resources required
 by these parasites and a discussion about host immune responses can be found in \cite{Telfer2008}.

\subsubsection{{\it Bartonella} species}
\label{sec: model bart}
{\it Bartonella} is a genus of bacteria that infects mammals (including humans), usually transmitted by arthropods. The species investigated 
here are transmitted by fleas (\cite{Bown2004}). We assume that the effect of other diseases and covariates on the probability that a
vole will recover from a particular {\it Bartonella} species after the
second (third fourth etc.) lunar month is the same as for the effect
on the probability of recovery after the first month; there are no
grounds for assuming otherwise. However, since the majority of {\it Bartonella} infections last for one month and only a few last more than 
two (\cite{Birtles2001}, \cite{Telfer2008}) the overall probabilities of recovering after the first and second month 
 are likely to be different. Additionally a vole's chance of contracting
 a particular {\it Bartonella} species for the first time might be different from the chance of contracting it again after recovery 
from it in the past, although again, there is no reason to assume that the effects of other diseases and covariates on this are likely to be
different. 
This suggests that each {\it Bartonella} species could be sensibly modelled using a Markov chain with four states: 1=no infection, 2=new 
infection, 3=old infection, 4=uninfected but has had a past infection. However, the observed sequence indicates either negative (N) or positive (P) status. In particular, an observation of $y^{[d]}_t=N$ corresponds to hidden process of $X^{[d]}_t=1$ or $X^{[d]}_t=4$ with likelihood vector ${\bf
  l}_t^{[d]}=(1,0,0,1)$, and 
an observation of $y^{[d]}_t=P$ corresponds to $X^{[d]}_t=2$ or
$X^{[d]}_t=3$ with ${\bf l}_t^{[d]}=(0,1,1,0)$. The time-inhomogeneous transition probability matrix from time $t$ to time 
$t+1$ for this Markov chain is 
\begin{eqnarray}
P = \left[ 
\begin{tabular}{c c c c}
$1- p_{12}$ & $p_{12}$ & 0 & 0 \\
0 & 0 & $1-p_{24}$ & $p_{24}$ \\ 
0 & 0 &$1-p_{34}$ & $p_{34}$ \\
0 & $p_{42}$ & 0 &$1-p_{42}$
\end{tabular} 
\right].
\end{eqnarray}
Each transition probability is governed by a logistic regression as follows:
\begin{eqnarray}
logit(p_{12}) &=& \beta_{0,12} + {\bf z}^{T}\boldsymbol {\beta}_{contract},  \\
logit(p_{24}) &= &\beta_{0,24} + {\bf z}^{T}\boldsymbol{\beta}_{recover}, \\
logit(p_{34}) &= &\beta_{0,34} + {\bf z}^{T}\boldsymbol{\beta}_{recover}, \\
logit(p_{42}) &= &\beta_{0,42} + {\bf z}^{T}\boldsymbol{\beta}_{contract}. 
\end{eqnarray}
As
justified above, we use 
the same vector of covariate effects $\boldsymbol{\beta}_{contract}$ for the two probabilities related 
to contracting the particular {\it Bartonella} species. Similarly we use the same covariate effects for the two probabilities 
relating to recovery from the disease, $\boldsymbol{\beta}_{recover}$; we allow
only the intercepts to differ.  
This assumption prevents a further increase in the, already large,
number of parameters to be estimated. For example, the logistic model
for the probability of contracting {\it B. taylorii} for the first
time at lunar month $\mbox{Lm}+1$ will be 
\begin{small}
\begin{eqnarray}
\label{eq:logistic}
\text{logit}(p_{12}^{tay}) &= & \beta_{0,12} + \beta_{wt}\text{weight}+\beta_{Lm}\text{Lm} +\beta_{sex}I(\text{male})+\beta_{s2}I(\text{site2})+\beta_{s3}I(\text{site3})+\beta_{s4}I(\text{site4}) \nonumber \\ 
&&+\beta_{sin}\sin +\beta_{cos}\cos+\beta_{sex:sin}I(male)\sin  +\beta_{sex:cos}I(\text{male})\cos + \beta_{grah2}I(\text{grah2}) \nonumber  \\ & &+ \beta_{grah3}I(\text{grah3})+ \beta_{grah4}I(\text{grah4}) +\beta_{dosh2}I(\text{dosh2})+ \beta_{dosh3}I(\text{dosh3})  \nonumber \\
&&+ \beta_{dosh4}I(\text{dosh4}) + \beta_{bab2}I(\text{bab2})+ \beta_{bab3}I(\text{bab3})+ \beta_{cow2}I(\text{cow2}) + \beta_{cow3}I(\text{cow3}) \nonumber \\ 
& &+ \beta_{ana2}I(\text{ana2}). \nonumber
\end{eqnarray}
\end{small}
Here and elsewhere $I(\cdot)$ denotes the indicator function, and
[disease]$x$ is a statement that the hidden chain for [disease] is
in state $x$. 

\subsubsection{\it Babesia}
\label{sec:babesia}
{\it Babesia microti} can cause haemolytic anaemia in infected
hosts. It is a chronic infection, which is to say that once a host
is infected it is never again free of the disease. The effect of a {\it Babesia}
infection on the probabilities of contracting or recovering from one
of the other diseases may depend on whether the {\it Babesia}
infection is acute (in its first month) or chronic.

We therefore model {\it Babesia} using a Markov chain with the following
three states: 
 1=no infection, 2=acute infection, 3=chronic infection. Here the
 likelihood vector that connects the states with the observations is analogous to that for {\it Bartonella} species but ignoring state 4. The transition
 matrix is
\begin{equation}
 P = \left[ 
\begin{tabular}{c c c }
$1- p_{12}$ & $p_{12}$ & 0 \\
0 & 0 & 1 \\ 
0 & 0 &1 \\
\end{tabular} \right]. \nonumber
\label{eq: babesia matrix}
\end{equation}
As in the previous section, a logistic regression relates $p_{12}$ to the covariates, including the states of the other diseases.

\subsubsection{\it Anaplasma}
{\it Anaplasma phagocytophilum} is a tick-borne bacterium that causes the disease granulocytic
ehrlichiosis in humans. In the dataset there are relatively few
positive records for {\it Anaplasma} and thus little power to
ascertain transition probabilities and covariate effects from a third
state of, for example, ``currently uninfected but was previously infected". Therefore, we use a two-state Markov chain with the following transition probability matrix

 $$P = \left[ 
\begin{tabular}{c c }
$1- p_{12}$ & $p_{12}$ \\
$p_{21}$ & $1-p_{21}$
\end{tabular} 
\right] ,$$
with separate logistic regressions relate $p_{12}$ and $p_{21}$ to the covariates, including
the states of the other diseases. This therefore is the only disease
for which the underlying Markov model is not hidden.

\subsubsection{Cowpox}

In voles and other wild rodents, infection with cowpox virus is known
to last for approximately 4 weeks (\cite{Bennett1997}). The 
diagnostic test, however, detects antibodies to the virus, not the
virus itself. Antibodies appear approximately 2 weeks after contracting the infection but then remain present in the blood stream of a vole 
for the rest of its life (\cite{Bennett1997}). 
Since the disease lasts for approximately one month we model the progression as a Markov chain with three
states: 1=antibodies absent and disease absent, 2=antibodies present and disease present,
3=antibodies present and disease absent. Therefore, the
form of the transition matrix and the relationship between the states and the
response is identical to that
for {\it Babesia}. The difference is in the interpretation: here State
3 corresponds to a positive response but absence of the disease, whereas for {\it
  Babesia} State 3 corresponds to a positive response which means that
the disease is present.

\section{Bayesian approach}

\subsection{Choice of prior}
\label{sec: prior}
\subsubsection{Initial probability distribution}
\label{sec: initial}

The likelihood (Section \ref{sec: likelihood}) and the forward-backward algorithm (Section \ref{sec: FW-BW}) require, for each disease, the initial distribution,
$\boldsymbol{\pi}^{[d]}$, of the Markov chain on the set of states for that disease at
the first observation time for each vole. 
Our
time-inhomogeneous Markov chains admit no limiting distribution and so
the popular choice of setting the initial distribution to the limiting
distribution of the chain is not available to us. 

We choose then to estimate this distribution for each disease through
our Gibbs sampler in Section \ref{sec: gibbs}. We choose independent
and relatively vague
Dirichlet priors,  $\bmpi^{[d]}\sim Dir(\boldsymbol{\alpha}^{[d]})$,
with  $\boldsymbol{\alpha}^{[d]}$ set to a vector of ones with length
equal to the cardinality of state space $\mathbb{S}^{[d]}$. This is
equivalent to a uniform prior on each $\bmpi^{[d]}$.

\subsubsection{Prior distributions for the regression parameters} 
\label{sec: priorpar}
A similar longitudinal dataset to the one that we analyse was also
available to us. This additional dataset arises from an earlier, 
three year study which was conducted using the same
 sampling design, but where the response for
{\it Bartonella} was a single indicator for presence and absence,
rather than an indicator for each species. For each of  
\textit{Babesia}, \textit{Anaplasma}, and cowpox we were therefore
able to fit a logistic regression to a subset of the additional dataset 
 as briefly described in Sections \ref{sec.motiv} and \ref{sec: model}, except that the
three indicator covariates for presence or absence of each \textit{Bartonella} species
were replaced with a single indicator covariate for presence or absence of at
least one
\textit{Bartonella} species. 
Parameter estimates
from these analyses were used to inform our choice of prior for
similar parameters in our main analysis. 

Since the additional dataset does not distinguish the {\it Bartonella}
species, there is not an exact correspondence between parameters from
the simple analyses and the parameters in our main model, and some of
the parameters in our main model have no counterpart in the simple
analyses. The priors for each $\bmbeta^{[d]}$ ($d=1,\dots,D$) in our main analysis are
independent and Gaussian with covariance matrices that we denote by $V^{[d]}$. For
\textit{Babesia}, \textit{Anaplasma}, and cowpox, 
where parameters do approximately correspond, we
set the prior mean for the parameter in our main analysis to the 
MLE for the corresponding parameter in the simple analysis of the
additional dataset and the block of $V^{[d]}$ associated with these
parameters to $9$ times the analogous block from the simple analysis. Where no corresponding MLE exists, the prior mean was set 
equal to zero and the block of $V^{[d]}$ was a diagonal matrix where the diagonal elements were set to nine.

\subsection{Adaptive Metropolis-within-Gibbs algorithm}
\label{sec: gibbs}

In our dataset, the target parameter, ${\boldsymbol \beta}$, can be
naturally partitioned into six sub-blocks, one for each disease. In the Gibbs scheme we wish to update the covariate parameters for each given disease $d$, $\bmbeta^{[d]}$, using a {\it random walk Metropolis} step (RWM) (see e.g. \cite{Gilks1996}); however the efficiency of a given RWM algorithm depends heavily on the choice
of the variance of the proposal jump, $\it \Sigma^{[d]}$.
Both \cite{Roberts2001} and
\cite{Sherlock2009} suggest that a RWM algorithm might achieve near
optimal efficiency when $ \it \Sigma^{[d]}$ correctly represents the general shape of the
target distribution, for example if it is proportional to the variance
of $\bmbeta^{[d]}$ or the inverse curvature at the mode. We therefore generalise the \textit{adaptive RWM} algorithm described in \cite{Sherlock2010} to an \textit{adaptive
Metropolis-within-Gibbs} algorithm on $D$ sub-blocks. 

Let ${\boldsymbol \beta}^{[1:D]}=({\boldsymbol \beta}^{[1]},{\boldsymbol \beta}^{[2]},\ldots,{\boldsymbol \beta}^{[D]}),$ where ${\boldsymbol \beta}^{[d]}$ 
is the parameter set for the $d$th sub-block. A single iteration starts from an initial value
 $\boldsymbol{\beta}^{[1:D]}$, cycles through all the sub-blocks updating each in
 turn, and finishes with 
${\boldsymbol \beta}'^{[1:D]}=({\boldsymbol \beta}'^{[1]},{\boldsymbol \beta}'^{[2]},\ldots,{\boldsymbol \beta}'^{[D]})$.
In the update for the $d^{th}$ block the current and proposed values
are respectively:
\begin{eqnarray}
\boldsymbol{\beta}^{[1:D]} & := & \boldsymbol{\beta}'^{[1]},\ldots,\boldsymbol{\beta}'^{[d-1]},\boldsymbol{\beta}^{[d]},\boldsymbol{\beta}^{[d+1]},\ldots,\boldsymbol{\beta}^{[D]}, \nonumber \\
\boldsymbol{\beta}^{*[1:D]}& := & \boldsymbol{\beta}'^{[1]},\ldots,\boldsymbol{\beta}'^{[d-1]},\boldsymbol{\beta}^{[d]}+\boldsymbol{\epsilon},\boldsymbol{\beta}^{[d+1]},\ldots,\boldsymbol{\beta}^{[D]}.
\label{eq:shorthand}
 \end{eqnarray}
Here the proposal jump $\boldsymbol{\epsilon}$ in (\ref{eq:shorthand}) for the $d$th
 sub-block at the $n$th iteration is sampled from a mixture distribution:
 \begin{equation} 
\boldsymbol{\epsilon}
\sim
\left\{
\begin{array}{ll}
N\left({\bf 0}, \left(m^{[d]}_n\right)^2 {\it \tilde{\Sigma}}^{[d]}_n\right)
& \text{with probability $1-\delta$,} \\
N\left({\bf 0}, \left(m^{[d]}_0\right)^2 {\it \Sigma}_0^{[d]}\right) & \text{with probability $\delta$.}
\end{array}
\right.
\label{eq:epsilon}
\end{equation}  
Here $\delta$ is a small positive constant, which we set to $0.1$,
 ${\it \Sigma_0^{[d]}}$ is a fixed covariance matrix and $m_0^{[d]}$ is set to the theoretically derived value $2.38/k^{1/2}$ (\cite{Roberts2001}), where $k$
is the dimension of $\bmbeta^{[d]}$. The matrix
 $\tilde{\it \Sigma}_n^{[d]}$ is the estimated variance of $\bmbeta^{[d]}$ using the
 sample from the Markov chain to date.
The scaling factor $m_n^{[d]}$ for 
the adaptive part is initialised to $m_0^{[d]}$. \cite{Sherlock2010} details the adaptation of the scaling factor at each iteration, which 
leads to an equilibrium acceptance rate of 25\%, close to the optimal
acceptance rate (approximately $23\%$) derived by
\cite{Roberts2001}. 

The conditional likelihood for $\bmbeta^{[d]}|{\bf y}^{[d]},{\bf
  x}^{[-d]},\bmpi^{[d]}$ is calculated using the forward part of the
forward-backward 
algorithm for disease $d$. This provides the conditional
posterior for $\bmbeta^{[d]}$ and hence the acceptance probability for the proposed value
of $\bmbeta^{[d]}$. The states of the hidden Markov
chain for disease $d$ can then be sampled from their conditional
distribution given the observed
data for that disease, $\bmpi^{[d]},~\bmbeta^{[d]}$, and the states of
the other diseases using the backwards part of the forward-backward algorithm.

Given the states of the hidden chain for a particular disease, $d$, and
in particular, the initial state for each vole, 
the conditional conjugacy of the Dirichlet distribution allows
straightforward sampling
 from the conditional posterior for $\bmpi^{[d]}$.

We therefore simulate from the joint posterior
distribution of the coefficients of the logistic regressions for the transition
 probabilities, the hidden disease states and the initial probability distribution of the hidden states with the following MCMC algorithm.

At the start of the current iteration of the chain let the 
 covariate parameters be
 ${\boldsymbol \beta^{[1:D]}}$, let the hidden states be ${\bf
   X}^{[1:D]}$ and the initial distribution of the hidden states be
 $\bmpi^{[1:D]}$; denote their values at the start of the next
 iteration as ${\boldsymbol \beta'^{[1:D]}}$, ${\bf X}'^{[1:D]}$ and $\bmpi'^{[1:D]}$ respectively.

Each step of the Gibbs sampler is as follows.
\begin{itemize}
\item Perform an adaptive RWM update according to $\boldsymbol{\beta}'^{[1]}|{\bf y}^{[1]},\boldsymbol{\beta}^{[1]},{\bf x}^{[2:D]},\bmpi^{[1]}$.
\item Simulate the hidden states for the first disease from ${\bf
    X}'^{[1]}| {\bf y}^{[1]}, {\boldsymbol \beta}'^{[1]},{\bf x}^{[2:D]},\bmpi^{[1]}$.
\item Simulate the initial probability distribution of the chain for the first disease $\boldsymbol{\pi}'^{[1]}|{\bf x}'^{[1]}$.
\item Perform an adaptive RWM update according to $\boldsymbol{\beta}'^{[2]}|{\bf y}^{[2]},\boldsymbol{\beta}^{[2]},{\bf x}'^{[1]},{\bf x}^{[3:D]},\bmpi^{[2]}$.
\item Simulate the hidden states for the second disease from 
${\bf X}'^{[2]}|{\bf y}^{[2]},{\boldsymbol \beta}'^{[2]},{\bf x}'^{[1]},{\bf x}^{[3:D]},\bmpi^{[2]}$. 
\item Simulate $\boldsymbol{\pi}'^{[2]}|{\bf x}'^{[2]}$.
\item $\ldots$
\item Perform an adaptive RWM update according to $\boldsymbol{\beta}'^{[D]}|{\bf y}^{[D]},\boldsymbol{\beta}^{[D]},{\bf x}'^{[1:D-1]},\bmpi^{[D]}$.
\item Simulate ${\bf X}'^{[D]}|{\bf y}^{[D]},{\boldsymbol \beta}'^{[D]},{\bf x}'^{[1:D-1]},\bmpi^{[D]}$ 
\item Simulate $\boldsymbol{\pi}'^{[D]}|{\bf x}'^{[D]}$.
\end{itemize}

The adaptive RWM step requires the fixed covariance matrix ${\it \Sigma^{[d]}_0}$. For each disease, a separate 
non-adaptive RWM was performed for the logistic regressions coefficients associated with the hidden Markov model for this disease that are not associated with the other diseases; for example weight 
and sex. The block of ${\it \Sigma^{[d]}_0}$ associated with these
covariates was estimated directly from this run. Each of the remaining
$\beta$ coefficients was given a small proposal variance  and was assumed to be uncorrelated with any of the other coefficients.
 Also to ensure a sensible non-singular $\hat{\it{\Sigma}}^{[d]}_n$, for each disease, proposals from the adaptive part were only allowed once at least 1000 
proposed jumps had been accepted. 

\section{Analysis and results}

\subsection{Convergence of the algorithm and model diagnostics}
All the computationally intensive parts of the algorithm were coded in
\texttt{C} within an \texttt{R} (\cite{R}) wrapper. On a computer with
an Intel Nehalem 2.26 GHz CPU, 100,000 iterations of the algorithm
took approximately 3 hours.

\begin{sidewaysfigure}
\centering
  \includegraphics[scale=0.71]{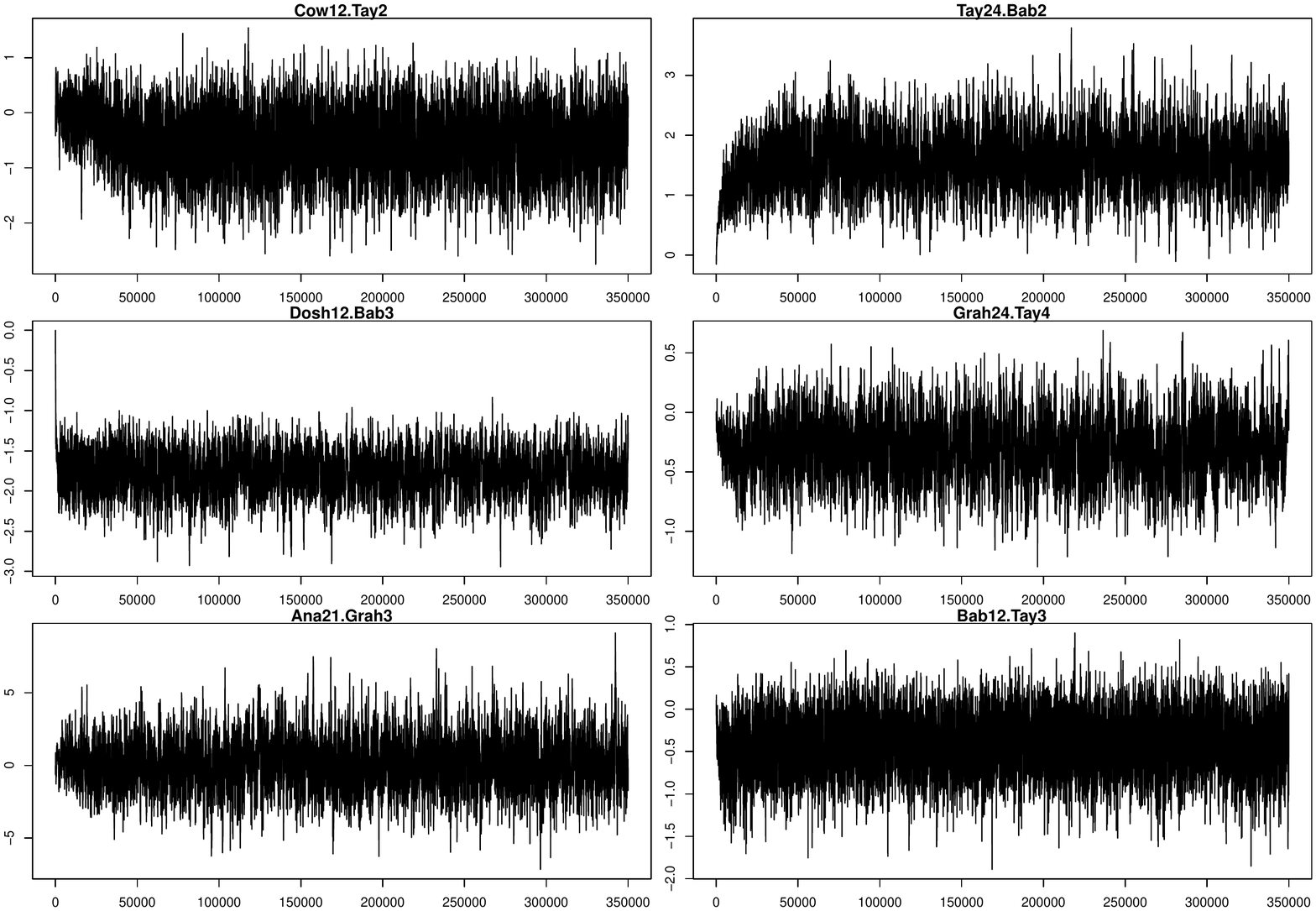}
\caption{\label{fig: traceplots}Six of the 233 traceplots for the first of the three runs of the Markov chain. Each 
trace plot represents the $\beta$ coefficient of the logistic regression for the transition probabilities of the HMM for a different disease.}
\end{sidewaysfigure}

Three independent Markov chains of length 350,000 were generated from the algorithm in Section \ref{sec: gibbs}; each chain was started from a different position.  
Six of the 233 trace plots from one of the chains are reproduced in
Figure \ref{fig: traceplots}. Most of these, over the first few 
tens of thousands of the iterations the variance of the proposal increases as the adaptive 
algorithm learns the shape of the posterior; this was the case in many of the 233 trace plots.

\begin{figure}[!ht]
  \centering
   \makebox{\includegraphics[scale=0.4]{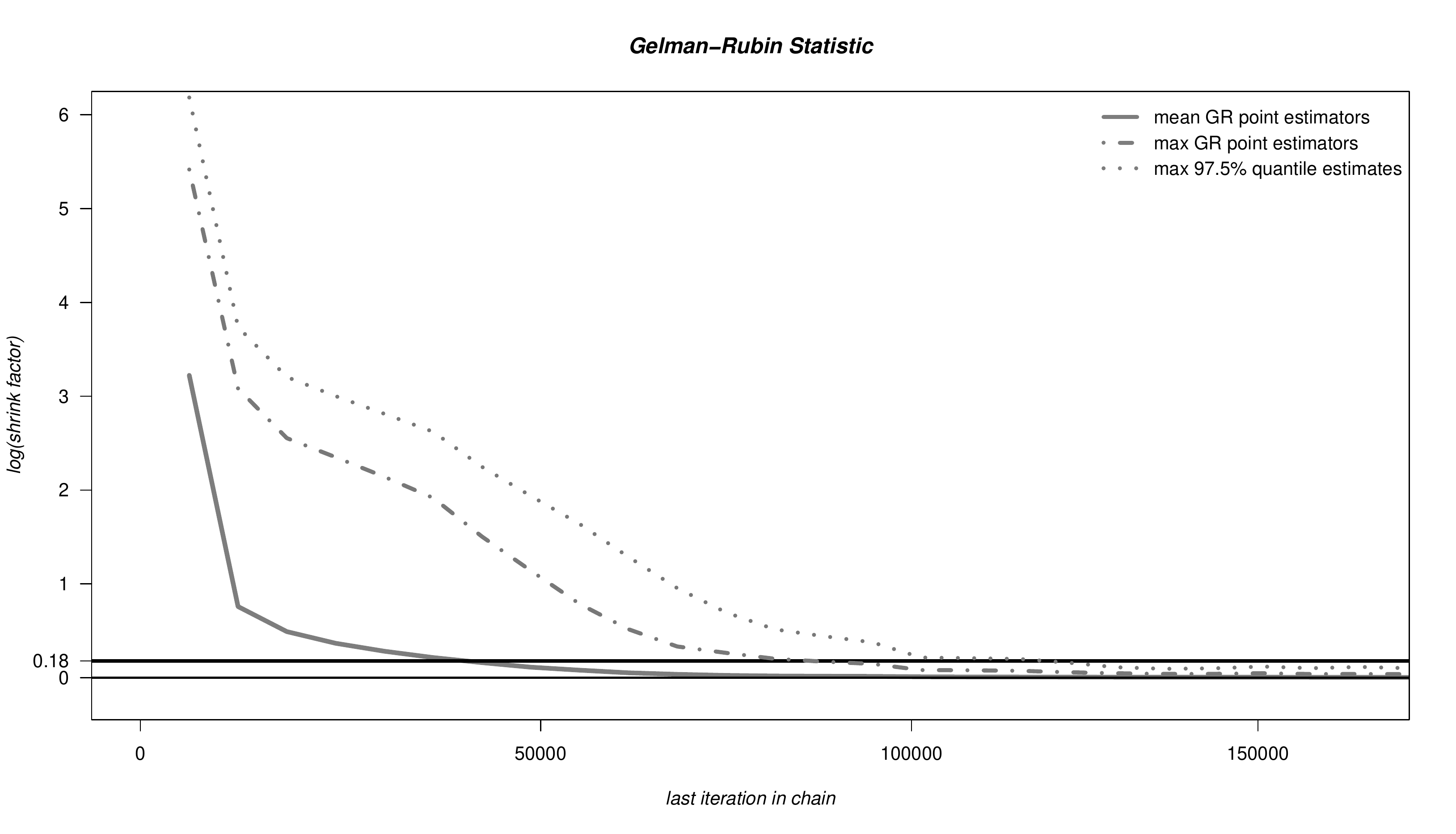}}
  \caption{\label{fig: gelman}Combined Gelman-Rubin statistics for all 233 $\beta$'s; $R_1,\dots,R_{233}$. $\frac{1}{233}\sum_iR_i$(---); $\max_i R_i$(-- -- --); $\max_i R_{i,0.975}$(- - -). Statistics are plotted on the log scale against iteration number, along with the ideal ratio of log(1) ({\bf ---}) and the threshold of log(1.2) (---) suggested in \cite{Gelman1996}.}
\end{figure}

The Gelman-Rubin statistic (\cite{Gelman1992}), $R$, was calculated from the three chains for each of the 233 components of $\bmbeta$. Figure \ref{fig: gelman} 
shows the mean of the estimated $R$ statistics, the maximum of the
estimated $R$ statistics along with maximum of the 97.5\% quantiles of
$R$ statistics, plotted against iteration number. The plot suggests
that a burn-in period of 150,000 iterations should be more than
sufficient. Inference is therefore performed using the final $200,000$ iterations from each of the 3 runs combined. 

To assess model fit we examine the posterior predictive distribution of the data
(see \cite{Robert1999}). We chose, at random, 100 captures where all
six diseases were observed and created an alternative dataset where
all diseases for these captures were marked as missing. 
We refitted the model and
estimated the posterior probabilities, $\hat{p}$, of these
artificially removed observations being positive. A Hosmer-Lemeshow
test (e.g. \cite{Collett2003}) for each disease 
provides a p-value for the null hypothesis that each of the true
(binary) observations 
arises from a Bernoulli trial with the given posterior
probability. For the six diseases we obtained p-values of  $0.443$
({\it B. doshiae}),  0.188 ({\it B. grahamii}), 0.061 ({\it
  B. taylorii}), $0.141$ ({\it Babesia}), 0.56 ({\it Anaplasma}) and
0.322 (cowpox).

\subsection{Posterior inference}
\label{sec: posterior}
We are interested in interactions between diseases, for example in whether or not presence or absence of disease $d_1$ affects the probability of a change of state for disease $d_2$. In each logistic regression for each transition matrix, we therefore examine the coefficients that correspond to the states of the other diseases. We are also interested (for \textit{Bartonella}) in whether or not the status of an infection (new or old) affects the chance of recovery, and in whether or not a previous infection affects the chance of a new infection with the same species; these correspond respectively to the contrasts: $\beta_{0,34}-\beta_{0,24}$ and $\beta_{0,42}-\beta_{0,12}$.  

Formal model choice, for example via reversible jump MCMC (\cite{Green}), is computationally infeasible here. Instead we take a high posterior probability that a given parameter or contrast is positive (or a high probability that it is negative) as indicating a 
potentially important effect. For an individual parameter we might consider $P(\mbox{positive})>0.975$ or $P(\mbox{positive})<0.025$ as indicating a 
likely effect. We are however interested in a total of $116$ parameters and contrasts which raises a similar problem to that of multiple testing in classical statistics. Whilst considering probabilities below $0.025$ or above $0.975$ to indicate a possible interaction, we therefore take probabilities below $0.00025$ and above $0.99975$ as indicating a very probable interaction.

Table \ref{tbl:posterior} shows those parameters for which the posterior probability of positivity is either above $0.975$ or below $0.025$.
The table shows the posterior median, a $95\%$ credibility interval, and the posterior probability that the parameter is positive.

Firstly, and most clearly, the presence of {\it Babesia} decreases the probability of contracting {\it Bartonella} and increases the probability of recovery from 
{\it Bartonella}.
This is true for both chronic and acute {\it Babesia} infections and for all three species of {\it Bartonella}. There is no evidence for the reverse interaction, that is
 for the presence 
of {\it Bartonella} affecting the chance of contracting {\it Babesia}. 

For two of the three {\it Bartonella} species ({\it B. taylorii} and
{\it B. grahamii}) it appears that a vole is less likely to be re-infected following previous exposure while it is more likely to recover from an old infection of {\it B. taylorii} than a 
new one. Furthermore, a vole that has recovered from a \textit{B. taylorii} infection is less likely to contract \textit{B. grahamii}.
In addition, there seems to be a decrease in the
probability of contracting {\it B. doshiae} when a vole has been exposed to {\it B. taylorii} whether or not it is still infected. Finally, infection with \textit{Anaplasma} appears to increase the probability of recovery 
from \textit{B. grahamii}; there was perhaps some evidence for the same
interaction with \textit{B. doshiae} with posterior probability $0.9593$. There is no evidence of a change to the probability of recovery 
from \textit{B. taylorii} (probability$=0.416$). It is also possible that a current infection with cowpox virus hinders recovery from \textit{B. doshiae} and previous exposure to cowpox prevents infection with {\it Babesia}.

\subsection{Sensitivity analysis}
\label{sec: sensitivity}

Three somewhat arbitrary choices were made in the set up of our model
and priors: the interpolation scheme that fills in missing weight
values, the prior for the initial 
distribution for the state of the hidden Markov model for each disease and vole, and the exact relationship between parameters estimated in the simple 
analysis of the alternative dataset and priors for parameters in the hidden Markov models for the main dataset.

An alternative for each of these choices is described below. For each alternative three further chains of length 350,000 were created and checked for 
convergence. Then any sizeable changes in the conclusions that would
be drawn from the posterior distributions of the parameters were
noted.

In the main analysis, missing weight values were filled in via linear interpolation. As an alternative we considered the logistic growth curve which was 
proposed in \cite{Burthe2009}. We assumed Gaussian residuals for the logarithm of weight and allowed the logistic growth parameters to depend on covariates such 
as the sex of the vole and the time of year; some of the coefficients were also allowed to include subject specific random effects. More details are provided 
in \cite{Xifara2012}. 

The initial distributions for the states of the
hidden Markov models for the diseases are assigned independent
Dirichlet priors with the parameter for each disease, $\boldsymbol{\alpha}^{[d]}$, a vector
of ones.  As an alternative prior we set each
$\boldsymbol{\alpha}^{[d]}$ to be a vector of twos.

In the main analysis, where there was a rough correspondence between a parameter in the simple analysis of the alternative dataset and one in the main analysis,
 we centered the Gaussian prior in the main analysis on the maximum likelihood estimate from the simple analysis and set the covariance matrix to be nine times 
the estimated covariance matrix from the simple analysis (Section \ref{sec: priorpar}). As an alternative we use vague but proper Gaussian priors for all parameters.

\begin{table}[!ht]
\caption{\label{tbl:posterior}Posterior summaries of model parameters
  of interest for which the posterior probability of positivity is either above 0.975 or below 0.025. 
For each parameter the posterior median, a $95\%$ credibility interval and the probability that the parameter is greater than zero is provided. 
Each parameter arises from a logistic regression coefficient for a particular transition probability in the hidden Markov model for a particular disease. 
The disease and transition appear in the first column, and second column indicates
 the particular disease and state that is influencing the transition probability. State $1$ is always taken to be the baseline. The contrasts $\beta_{0,42}-\beta_{0,12}$ and $\beta_{0,34}-\beta_{0,24}$ are defined in Section \ref{sec: posterior}. }
 \centering
\fbox{%
\begin{tabular}{l l c c c c c c c c c} 
\em{Transition} &&\em{Covariate} & &\em{Median} & &&\em{$95\%$ CI} &&& \em{Posterior } \\
\em{Probability}&& &  & & & &  &&&\em{Probability}  \\ \hline
\multirow{5}{*}{$p^{dosh}_{12}$}&&tay2&& -1.0124 &&& (-1.8708, -0.1554) &&&0.0111\\ 
&&tay3&&-1.1077&&&(-2.0584, -0.1877)&&&0.0096\\ 
&&tay4&&-1.0304&&&(-1.8046, -0.2278)&&&0.0077\\
&&bab2&&-1.5636&&&(-2.4247, -0.8039)&&&0.0001\\
&&bab3&&-1.7702&&&(-2.3095, -1.3235)&&&0.0000\\ \hline
\multirow{4}{*}{$p^{dosh}_{24}$}&&bab2&&2.9613&&&(1.6459, 4.5298)&&&1.0000\\
&&bab3&&3.3386&&&(1.775, 5.5524)&&&1.0000\\
&&cow2&&-1.01&&&(-2.0937, -0.0691)&&&0.0171\\ \hline
\multirow{2}{*}{$p^{grah}_{12}$}&&bab2&&-1.2579&&&(-2.0506, -0.5881)&&&0.0000\\
&&bab3&&-2.0294&&&(-2.6841, -1.5096)&&&0.0000\\ \hline
\multirow{3}{*}{$p^{grah}_{24}$}&&bab2&&2.7818&&&(1.1981, 5.0649)&&&0.9999\\
&&bab3&&1.7288&&&(0.8899, 2.754)&&&1.0000\\
&&ana2&&1.0284&&&(0.0116, 2.1709)&&&0.9764\\ \hline
\multirow{2}{*}{$p^{tay}_{12}$}&&bab2&&-1.6933&&&(-2.6563, -0.815)&&&0.0001\\ 
&&bab3&&-1.3346&&&(-1.8338, -0.8754)&&&0.0000\\ \hline
\multirow{3}{*}{$p^{tay}_{24}$}&&bab2&&1.5824&&&(0.702, 2.5404)&&&0.9997\\
&&bab3&&1.7513&&&(1.0015, 2.6279)&&&1.0000\\ \hline
$p^{bab}_{12}$ && cow3 &&-0.4939&&&(-0.9799, -0.0206)&&&0.02 \\ \hline
\multicolumn{5}{l}{\em{Contrasts}} &&&& \\
\cline{1-4}
\multicolumn{4}{l}{$\beta^{grah}_{0,42}-\beta^{grah}_{0,12}$}&-1.3248&&&(-5.6452, -0.2774)&&&0.0076\\
\multicolumn{4}{l}{$\beta^{tay}_{0,42}-\beta^{tay}_{0,12}$}&-4.472&&&(-8.6604, -2.1214)&&&0.0015\\
\multicolumn{4}{l}{$\beta^{tay}_{0,34}-\beta^{tay}_{0,24}$}&1.1762&&&(0.555, 1.8601)&&&0.9998\\
\end{tabular}
}
\end{table}

Parameter estimates with the alternative weight scheme or with the alternative prior distribution of the hidden states were very similar to the estimates from the main set-up. For all the significant covariates none of the
posterior probabilities changed by more than $0.005$.
However, the use of vague priors for the parameters noticeably affected one 
of the twenty one covariates in Table \ref{tbl:posterior}. The effect on a vole's
probability of contracting {\it Babesia} when the vole had been exposed to cowpox became
apparently unimportant, with posterior probability changing from $0.02$ to $0.093$. 
No additional covariates became potentially important (i.e. $p<0.025$
or $p>0.975$) in any of the three alternative runs. 

\section{Discussion}
\label{sec:discussion}
We have described a coupled discrete-time hidden Markov model for interactions between diseases in a host and used it to analyse data from a longitudinal study of field
 voles with records of six different pathogens. The Markov model
 offers a more detailed description than the existing modelling
 approach that is described in Section \ref{sec.motiv}. Furthermore, by explicitly dealing with the
  missing observations (which comprise approximately $50\%$ of the dataset), the inference methodology that we introduce is able to use more 
of the data than the existing standard inference methodology.

Inference is performed via a Metropolis-within-Gibbs sampler that cycles
through the diseases and, for each disease conditional on the hidden
states of all of the
other diseases, samples from the
 parameters of the logistic regressions for the transition probabilities of the hidden Markov
 model using an adaptive random walk Metropolis step and then from the exact distribution
 of the hidden states given these parameters. These two steps use respectively the forwards and backwards
 parts of the forward-backward algorithm (FB).
 
The FB
 Gibbs sampler (e.g. \cite{Chib1996}, \cite{Scott2002} and
 \cite{Fearnhead2006}) also uses the forward-backward algorithm; however the motivation is
 different. The FB Gibbs sampler does not use the likelihood from the
 forwards recursion directly, as this would require a
 Metropolis-Hasting (MH) update; instead the backwards recursion provides
 a sample from the posterior distribution of the hidden states given
 the parameters. Due to the conditional conjugacy structure of the
 problems targeted by the FB Gibbs sampler it is then possible to sample exactly from the
 conditional posterior for the parameters given the hidden states and
 thus avoid the MH step and the associated tuning. Our logistic regression
 model for the transition probabilities does not allow a simple Gibbs
 step for updating the parameters conditional on the hidden states,
 and so we content ourselves with a MH step for the parameters and,
 for efficiency of mixing, do not condition on the hidden states for
 the current disease. After the MH step we \textit{then} sample from the
 hidden states for the disease so that these can be used as covariates
 for the other diseases; in effect, we therefore sample from the joint
 conditional distribution of the parameters and the hidden states for
 the disease. The
 FB algorithm could be avoided entirely by updating the logistic
 regression parameters conditional 
on the hidden states for \textit{all} diseases,
 and by sampling from the distribution of each individual hidden state conditional on all of the other hidden
 states and the transition parameters (e.g. \cite{RobertCe},
 \cite{RobertTit}). However we believe that the correlation between hidden states and
 between these states and the parameters would have led to a very
 poorly mixing MCMC chain; \cite{Scott2002} discusses the
 first aspect of this. 

\cite{Brand1997}, \cite{Jordan1999}, \cite{Rezek2001}, \cite{GhoshZhong2002} and \cite{NatarajanNevatia:2007} all
examine inference for coupled HMMs. The DAG for the HMMs considered in
these articles is the same as in Figure \ref{fig: HMM}, however in
order to allow recursions similar to those in the forward-backward algorithm all of
these articles - except \cite{Rezek2001} - make the simplifying assumption that the transition
probability for a given chain conditional on the others is separable:
\[
P\left(x_{t+1}^{d}|{\bf x}_t^{[1:D]}\right)
\propto
\prod_{i=1}^Df_{i,d}\left(x_{t+1}^{d}|x_t^{[i]}\right),
\]
for some collection of non-negative functions $f_{i,d}$.
Moreover the computational complexity increases with the square of the
sum of the  number of states in each chain, in practice each article considers only two chains.
We
wish to apply logistic regression rather than assume separability and to consider six chains; furthermore computational complexity of our algorithm increases with the sum of
the squares of the number of states in each chain. \cite{Rezek2001} perform Bayesian inference via a Gibbs sampling algorithm that is qualitatively similar to our own; the fully conjugate structure that is used does not, however, allow for the effects of any covariates.

We now examine the most major findings of Section \ref{sec: posterior}
and briefly discuss the biological insights that they offer. For {\it
  B. taylorii}, voles are more likely to recover from
an old infection than a new one, which is to be expected given that
more complete histories for individuals indicate that most infections
last for only one month. Previous data also indicate that {\it
  B. doshiae} infections may last longer than {\it B. taylorii} and
{\it B. grahamii} infections (\cite{Telfer2007}). Also, for {\it
  B. grahamii} and {\it B. taylorii}, 
previous infection by the species appears to grant some form of
immunity to that species, suggesting that hosts can develop an
effective acquired immune response. To date, there has been
conflicting evidence for such a response in wild populations,
suggesting immune responses may vary between host species and/or {\it
  Bartonella} species (\cite{Birtles2001}, \cite{Kosoy2004},
\cite{Bai2011}). 
Interestingly also {\it B. taylorii} infection appears to provide
immune cross-protection to {\it B. doshiae} infection. 

We found, for voles currently infected with {\it Babesia}, both a
reduction in susceptibility to {\it Bartonella} and an increase in the
probability of recovery from {\it Bartonella} over the next lunar
month. We also found no evidence that a current
{\it Bartonella} infection might influence susceptibility to {\it
  Babesia} over the next month. 
\cite{Telfer2010}
find the {\it Babesia} covariates, both at time $t_0$ and $t_1$, to be
significant for predicting the probability of catching {\it
  Bartonella} between $t_0$ and $t_1$. The {\it Bartonella} covariate
at time $t_1$ is also found to be significant in predicting
susceptibility to {\it Babesia}, apparently contradicting our
findings.
However, as mentioned in Section \ref{sec: model},
 \cite{Telfer2010} allow both the
current ($t_0$) and \underline{future ($t_1$)} state of each disease
covariate to
influence the probability that, for the disease that is being treated
as a response, a vole is positive at time $t_1$ given that it is
negative at time $t_0$. An assumption of only the negative effects of {\it
  Babesia} on {\it Bartonella} infections apparent from inference for
our HMM, and
no other dependency between {\it Bartonella} and {\it Babesia}, is
sufficient to lead to a negative correlation between {\it Bartonella}
and {\it Babesia} at any given time. Consider two groups of voles,
those with {\it Babesia} at $t_0$ (Group A) and those without {\it Babesia} at
$t_0$ (Group B). Since {\it Babesia} is a chronic infection, Group A
voles are more likely to have {\it Babesia} at  $t_1$ than voles from Group B. However
since {\it Babesia} impacts negatively on the probability of a vole
having {\it Bartonella}, Group A voles are less likely to have {\it
  Bartonella} at $t_1$ than voles from Group B.  Since the effect of
{\it Babesia} on {\it Bartonella} is so pronounced (Table \ref{tbl:posterior}), it is
certainly believable that this negative correlation could be strong
enough that each of the two diseases at $t_1$ appears as an important
covariate for the other. 

In the application which we have considered, missingness was
believed to be independent of disease state; in other scenarios, such
as those considered in \cite{Pradel2005} the probability that a given subject will be observed might depend on the states of each of the hidden Markov models. This could be accommodated within our methodology through a further logistic regression for the probability of being observed given the set of hidden states and other covariate information, and several other minor changes as detailed in \cite{Pradel2005}.

\section*{Acknowledgements}
Part of this work was funded through North West Development Agency project number N0003235.
T.X. acknowledges also funding from the EPSRC and FST at Lancaster University. The field work was supported by funding from the Natural Environment
Research Council (GR3/13051) and The Wellcome Trust (075202/Z/04/Z; 070675/Z/03/Z).

\bibliographystyle{Chicago} 

\end{document}